# Search of low-dimensional magnetics on the basis of structural data: spin-1/2 antiferromagnetic zigzag chain compounds $In_2VO_5$, $\beta−Sr(VOAsO_4)_2$, $(NH_4,K)_2VOF_4$ and $\alpha−ZnV_3O_8$

**L M Volkova**

Institute of Chemistry, Far Eastern Branch of the Russian Academy of Sciences
690022 Vladivostok, Russia

E-mail: volkova@ich.dvo.ru (L M Volkova)

**Abstract**
A new technique for searching low-dimensional compounds on the basis of structural data
is presented. The sign and strength of all magnetic couplings at distances up to 12 Å in five
predicted new antiferromagnetic zigzag spin-1/2 chain compounds $In_2VO_5$,
$\beta−Sr(VOAsO_4)_2$, $(NH_4)_2VOF_4$, $K_2VOF_4$ and $\alpha−ZnV_3O_8$ were calculated. It was stated that
in the compound $In_2VO_5$ zigzag spin chains are frustrated, since the ratio $(\alpha=J_2/J_1)$ of
competing antiferromagnetic (AF) nearest- $(J_1)$ and AF next-to-nearest-neighbour $(J_2)$
couplings is equal to 1.68 that exceeds the Majumdar-Ghosh point by 1/2. In other
compounds the zigzag spin chains are AF magnetically ordered single chains as $\alpha \rightarrow 0$. The
interchain couplings were analyzed in detail.

## 1. Introduction

The discovery of new spin-dependent phenomena has promoted the interest in low-dimensional
magnetic compounds. Special attention is drawn to compounds containing spin dimmers, linear and
zigzag spin chains, spin chains interacting through 'auxiliary' magnetic ions and spin ladder
systems as well as ordinary and double spin planes. Among the recent works on low-dimensional
spin systems one should mention the intensive studies of the effect of geometric frustration on
square, triangular and Kagome lattices as well as in one-dimensional quantum spin systems such as
zigzag chains and zigzag ladders.

In view of the above the development of new methods for the search and study of low-
dimensional magnetics becomes crucially important not only theoretically, but also in practical
terms of creating new materials with unique magnetic properties. The search of new materials with
desired magnetic structure can be performed on the basis of determining the magnetic coupling
parameters and magnetic structure from the compound crystalline structure data. For this purpose,
one can use an enormous bulk of materials on crystalline structures of magnetic compounds
available in the Inorganic Crystal Structure Database (ICSD) (FIZ Karlsruhe, Germany).

The main problem in this regard is that spatial location of magnetic ions in a crystalline
structure and the compound magnetic structure do not always coincide. In many compounds the
nearest magnetic couplings are weaker than longer-distance couplings inside low-dimensional
structural fragments or between them. However, we managed to reveal [1] the dependence of the magnetic couplings strength and the type of magnetic moments ordering on the relation between several crystal chemical parameters: (a) geometrical location of intermediate ions in local space between magnetic ions; (b) intermediate ions sizes; (c) distance between magnetic ions. This dependence is a rough model of generally known conceptions [2-4] on the determination of magnetic coupling parameters by electron shells overlapping.

On the basis of this dependence we developed a new phenomenological method [1] to estimate quantitatively the magnetic coupling parameters from the structural data of low-dimensional crystalline compounds. This method was named as 'crystal chemical method'. In spite of a rough character of the model, our method provides reasonable estimations not only on the spins orientation, but also on the strength of the whole spectrum of magnetic couplings as inside the low-dimensional fragment as between the fragments. The method is sensitive to slight changes in the magnetic ion's local environment. Use of this method to determine the magnetic coupling parameters enables one, in combination with analysis of magnetic coupling competition at specific geometrical configurations, to state the substance's magnetic structure on the basis of structural data.

In this paper we show the application of the crystal chemical method to search low-dimensional magnetics among compounds of known structures (section 2) and present a detailed study of magnetic couplings in five newly revealed magnetics $In_2VO_5$ [5], $\beta-Sr(VOAsO_4)_2$ [6], $(NH_4)_2VOF_4$ [7], $K_2VOF_4$ [8] and $\alpha-ZnV_3O_8$ [9] (section 3).

## 2. Technique

The sign and strength of magnetic couplings in compounds were calculated by a new crystal chemical method [1] on the basis of structural data with using the program "MagInter". The initial data format for the program "MagInter" (crystallographic parameters, atom coordinates) correspond to cif-file of the database (ICSD). The room-temperature structural data and ionic radii (IR, CN=6) of Shannon [10] ($r_{V^{4+}} = 0.58$ Å, $r_{O^{2-}} = 1.40$ Å, $r_{F^{1-}} = 1.33$ Å, $r_{N^{3-}} = 1.46$ Å, $r_{As^{5+}} = 0.46$ Å) were used for calculations.

One should mention that use of room-temperature structural data for calculations of couplings at low temperature could produce errors mainly in cases when intermediate ions are located in critical positions and slight deviations from these positions results in dramatic changes of the coupling strength or emerging of a phase transition of the 'antiferromagnetic (AF) – ferromagnetic (FM)' type. Besides, significant errors can be produced by deviations from ideal compositions and structural disorders of real crystals.

The following studies were consecutively performed in this work:

- 40 stoichiometric oxygen-containing compounds of V(IV) with distances between $V^{4+}$ ions in the structure in the range 2.8 – 4 Å were selected from the Inorganic Crystal Structure Database (ICSD);
- the sign and strength of all magnetic couplings between $V^{4+}$ at distances up to ~12 Å were calculated, and the spatial location of these couplings on the sublattice of magnetic ions in a crystal was stated;
- the probability of the emergence of anomalous magnetic couplings and magnetic phase transitions at insignificant changes in local space between magnetic ions was determined;
- specific geometrical configurations in magnetic ions sublattices hosting the magnetic couplings competition were identified;
- the conclusion on the compound's magnetic structures was made on the basis of obtained data on magnetic coupling parameters and availability of these coupling's geometrical frustrations.

As a result, it was found that the majority of the 40 selected compounds are low-dimensional magnetics. A detailed analysis of magnetic couplings and geometrical frustrations in seven of these compounds - $(Zn_2(VO)(PO_4))$, $(VO)(H_2PO_4)_2$, $(VO)SiP_2O_8$, $(VO)SO_4$, $(VO)MoO_4$, $Li_2(VO)SiO_4$ and $Li_2(VO)GeO_4$ - containing spin-1/2 linear chains and square lattices was presented in [11]. Magnetic properties of three of the compounds - $(VO)(H_2PO_4)_2$, $(VO)SiP_2O_8$ and $(VO)SO_4$ – were not studied previously. Among the predicted low-dimensional magnetics, only in five of them - $In_2VO_5$, $\beta-Sr(VOAsO_4)_2$, $(NH_4)_2VOF_4$, $K_2VOF_4$ and $\alpha-ZnV_3O_8$ – the magnetic structure is formed by AF zigzag spin-1/2 chains. Parameters of the main magnetic couplings in these zigzag chain antiferromagnets are presented in the table 1.

**Table**. Parameters of main intra- and interchain couplings calculated on the basis of the structural data.

|  | $In_2VO_5$ | $\beta-Sr(VOAsO_4)_2$ | | $(NH_4)_2VOF_4$ | $K_2VOF_4$ | $\alpha-ZnV_3O_8$ |
|---|---|---|---|---|---|---|
| Intrachain |  |  |  |  |  |  |
| d(V-V) (Å) | 3.268 | 3.535 | 3.511 | 4.196 | 4.193 | 5.676 |
| $J_{NN}^s$ (Å$^{-1}$) [a] | -0.085 AF | -0.156 AF | 0.153 AF | -0.135 AF | -0.129 AF | -0.070 AF |
| d(V-V) (Å) | 3.468 | 4.927 | 4.927 | 5.770 | 5.650 | 9.908 |
| $J_{NNN}^s$ (Å$^{-1}$) [b] | -0.143 AF | 0.006 FM | 0.003 FM | 0 | 0 | 0.001 FM |
| $J_{NNN}^s / J_{NN}^s$ | 1.68 | -0.04 | -0.02 | 0 | 0 | -0.01 |
| Interchain |  |  |  |  |  |  |
| d(V-V) (Å) | 10.484 | 8.344 (V1-V2) | | 7.596 | 9.313 | 7.340 |
| $J_{\text{int}erchain}^{s(\max)}$ (Å$^{-1}$) [c] | -0.040 AF | -0.033 AF | | -0.034 AF | -0.027 AF | -0.032 |
| $J_{\text{int}erchain}^{s(\max)} / J_{NN}^s$ | 0.47 | 0.21 | | 0.25 | 0.21 | 0.46 |
| $J_{\text{int}erchain}^{s(\max)} / J_{NNN}^s$ | 0.28 |  |  |  |  |  |

[a] Nearest-neighbor intrachain coupling.
[b] Next-to-nearest-neighbor intrachain coupling.
[c] Maximum interchain coupling.

## 3. Results and discussion

### 3.1. $In_2VO_5$

The compound $In_2VO_5$ [5] crystallizes in the Pnma system with $a = 7.232$ Å, $b = 3.468$ Å, and $c = 14.82$ Å. The $VO_6$ coordination polyhedron is a distorted octahedron with short vanadyl bond V-O1 (1.76 Å), substantially elongated bond V-O5 (2.23 Å) located in trans-position to it and four bonds at distances 1.82 - 2.03 Å in the equatorial plane. The bond-valence sum of V ions (BVS=4.24), calculated according to [12], slightly exceeds the ideal value. These octahedra share an edge (O5-O5) and form a zigzag chain parallel to the $b$ axis (figure 1(a)). In the zigzag chain the bond angles of V–O5–V are equal to 107° (edge sharing) and 146° (corner sharing), and the angle V-V-V (bend angle) is equal to 64.1° (figure 1 (a)-(c)).

The magnetic $V^{4+}$ ions carry S = 1/2 and form zigzag chains along the $b$ axis with strong AF nearest-neighbor $J_1$ ($J_1^s$  -0.085 Å$^{-1}$, d(V-V) = 3.268 Å) and even stronger competing AF next-to-nearest-neighbor $J_2$ ($J_2^s$ = -0.143 Å$^{-1}$, d(V-V) = 3.468 Å) couplings (figure 1 (d)-(f)). The zigzag chain can be also presented as a triangular two-leg ladder consisting of two linear chains ($J_2$ coupling along legs) with zigzag coupling ($J_1$ coupling along rungs).

The nearest-neighbor $J_1$ coupling forms under effect of two intermediate oxygen ions O5 localized in the central one-third part of the space ($l = 1.376$ Å and $l' = 1.892$ Å) between ions $V^{4+}$ at distance 1.184 Å ($h(O_5)$) from the centre of O5 ion to the straight line V-V connecting magnetic ions $V^{4+}$ (figure 1(b)). Every O5 ion contributes ($j_{O5}^s$ = -0.0424 Å$^{-1}$) to emerging of the AF component of the $J_1$ coupling. The next-to-nearest neighbour $J_2$ coupling emerges under effect of only one O5 ion, however, its contribution ($j_{O5}^s$ = -0.1434 Å$^{-1}$) to the AF-component of this coupling is much larger than the contribution from two O5 ions in the case of $J_1$ coupling, since it is located closer to the line V-V ($h(O_5)$ = 0.538 Å; $l'/l$ = 1) (figure 1(c)).

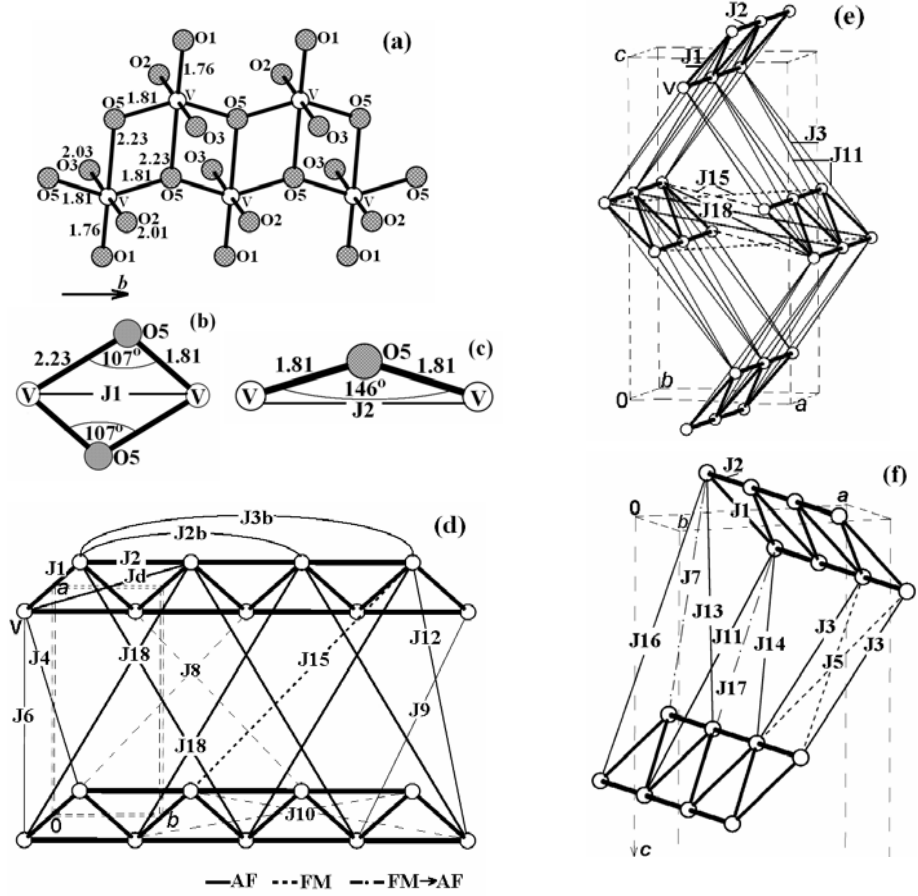

**Figure 1.** (a) Zigzag chain formed by edge-sharing VO$_6$ octahedra in In$_2$VO$_5$. The arrangement of intermediate ions in space of $J_1$ (b) and $J_2$ (c) couplings. The sublattice of V$^{4+}$ and coupling $J_n$ in In$_2$VO$_5$: (d) $ab$ plane, (e) three-dimensional structure and (f) couplings between zigzag chains from neighbouring planes. In this and other figures the thickness of lines shows the strength of $J_n$ coupling. AF and FM couplings are indicated by solid and dashed lines, respectively. The possible FM→AF transitions are shown by stroke in dashed lines.

In order to confirm or disprove the existence of competition between main couplings $J_1$ and $J_2$ and additional couplings at long distances in the zigzag chain, we calculated the parameters of four more magnetic couplings. Two couplings in linear chains along the $b$ axis (along legs) $J_{2b}$ ($J_{2b}/J_2$ = 0.15, d(V-V) = 6.936 Å = 2$b$) and $J_{3b}$ ($J_{3b}/J_2$ = 0.15, d(V-V)= 10.404 Å = 3$b$) are antiferromagnetic (figure 1(d). Competition between the nearest $J_2$ and next-to-nearest-neighbour $J_{2b}$ couplings can not induce a spontaneous dimerisation in linear chains, since the ratio $J_{2b}/J_2$ = 0.15 is less than the critical value as 1/6 in [13] and 0.2411 in [14 - 16]. However, at slight displacements of intermediate O1 and O5 oxygen ions under effect of temperature the coupling strength $J_{2b}$ can increase, and the value of $J_{2b}/J_2$ can attain the critical value. One of two additional couplings between linear chains $J_d$ ($J_d^s/J_2^s$ = 0.16, d(V-V) = 5.893 Å) (figure 1(d)) is antiferromagnetic and forms (with $J_1$ and $J_2$ couplings) an AF-triangle with competing unequal couplings. Another coupling $J_{10}$ at longer distance (d(V-V) = 9.102 Å) is a very weak ferromagnetic one ($J_{10}^s/J_2^s$ = -0.01).

Couplings between zigzag chains located in planes parallel to the *ab* plane (figure 1(d)) through parameter *a* are stronger than those between chains from neighbouring planes (figure 1(f)). The strongest couplings are those at long distances: AF $J_{18}$ coupling ($J_{18}^s/J_1^s$ = 0.47, $J_{18}^s/J_2^s$ = 0.28, d(V-V) = 10.484 Å) and FM $J_{15}$ coupling ($J_{15}^s/J_2^s$ = -0.19, d(V-V) = 10.020 Å). There are no strong couplings between chains from neighbouring planes *ab*; the maximum one among them - the AF $J_3$ coupling - is two times weaker than the $J_{18}$ coupling.

All the interchain couplings at short distances as in the *ab* plane ($J_4$ (AF, $J_4^s/J_2^s$ = 0.10, d(V-V) = 6.332 Å) and $J_6$ (AF, $J_6^s/J_2^s$ = 0.008, d(V-V) = 7.232 Å = *a*)) as between planes ($J_3$ (AF, $J_3^s/J_2^s$ = 0.13, d(V-V) = 6.282 Å) and $J_5$ (FM, $J_5^s/J_2^s$ = -0.09, d(V-V) = 7.175 Å)) are also weak. The ratio $J_n^s/J_2^s$ for weak AF interchain couplings at long distances $J_{11}$ (d(V-V) = 9.216 Å), $J_{12}$ (d(V-V) = 9.266 Å), $J_{13}$ (d(V-V) = 9.279 Å), $J_{14}$ (d(V-V) = 9.358 Å) and $J_{16}$ (d(V-V) = 10.337 Å) fall inside the range 0.07-0.12. One should mention that two weak AF $J_7$ ($J_7^s/J_2^s$ = 0.004, d(V-V) = 7.877 Å) and FM $J_{17}$ (, $J_{17}^s/J_2^s$ = -0.04, d(V-V) = 10.440 Å) couplings between chains from neighbouring planes could change the type of spin orientation without the coupling strength change.

Thus, the magnetic structure of $In_2VO_5$ comprises AF frustrated zigzag spin chains along the *b* axis. The ratio of the AF nearest- $J_1$ and AF next-to-nearest-neighbor couplings $J_2$ in zigzag chains ($\alpha = J_2/J_1$ = 1.68) exceeds the Majumdar-Ghosh point ($\alpha$ = 1/2) [17] at which the ground state consists of dimerized singlets with a gap to the excited states. There exist different opinions in regard to the ground state of the $J_1$-$J_2$ Heisenberg chain for a large range around this point. It was shown in [18, 19] that in the extreme case $J_2 \gg J_1$ both dimerization and incommensurate spiral spin correlations are exhibited, whereas, according to [20], the incommensurate ground state is absent for large $J_2$. Relatively strong AF $J_{18}$ couplings ($J_{18}^s/J_1^s$ = 0.47, $J_{18}^s/J_2^s$ = 0.28) exist between zigzag chains in the plane *ab* at long distances. These couplings are two times stronger than maximum couplings ($J_3$) between chains from neighbouring planes.

### 3.2. ß-Sr(VOAsO₄)₂

The compound ß-Sr(VOAsO₄)₂ [6] crystallizes in the orthorhombic space group P2₁2₁2₁ with *a* = 4.927 Å, *b* = 12.565 Å, and *c* = 12.739 Å. The vanadium ions occupy two crystallographically independent sites V1 and V2 and have octahedral environment with short vanadyl bond (V1-O9=1.65 Å, V2-O10=1.63 Å) and five V-O bonds in the range 1.95 -2.09 Å. The bond-valence sum for V1 (4.13) and V2 (4.22) slightly exceeds the formal oxidation state $V^{4+}$. In the structure of ß-Sr(VOAsO₄)₂ one should mention two types - [V(1)O₄O₂/₂]∞ and [V(2)O₄O₂/₂]∞ - of infinite zigzag chains of corner-sharing VO₆ octahedra running in the same direction along the *a*-axis (figures 2(a), (b)). In both chains short V=O (~1.6 Å) bonds alternate with long V-O (2.0 Å) bonds while valent angles V1-O-V1(V2-O-V2) and angles between vanadium ions V1-V1-V1 (V2-V2-V2) are equal to 151.5(151.9)° and 88.4(89.1)°, respectively.

The magnetic and crystalline structures of the magnetic ions sublattice in the compound ß-Sr(VOAsO₄)₂ coincide. Let us denote magnetic nearest- and next-to-nearest-neighbor couplings as $J_1^{(1)}$ and $J_2^{(1)}$ in zigzag chain of V1 ions (figure 2(c)) and as $J_1^{(2)}$ and $J_2^{(2)}$ in zigzag chain of V2 ions (figure 2(d)), respectively. Parameters of respective magnetic interactions are virtually equal in both chains. The nearest-to-neighbor couplings $J_1^{(1)}$ ($J_1^{s(1)}$ = -0.156 Å⁻¹ (AF), d(V1-V1) = 3.535 Å) and $J_1^{(2)}$ ($J_1^{s(2)}$ = -0.153 Å⁻¹ (AF), d(V2-V2) = 3.511 Å) are dominating AF- couplings in ß-Sr(VOAsO₄)₂. One should note that a substantial contribution to the AF-component of the nearest-neighbor couplings $J_1^{(1)}$ emerges under effect of the intermediate ion O9 ($j_{O9}^s$ = -0.156 Å⁻¹; $h$(O9) = 0.445 Å, $l'/l$ = 1.23), while $J_1^{(2)}$ - under effect of the ion O10 ($j_{O10}^s$ = -0.160 Å⁻¹; $h$(O10)= 0.435 Å, $l'/l$ = 1.24). However, the value for the $J_1^{(2)}$ coupling slightly reduces due to a small contribution ($j_{O4}^s$ = 0.007 Å⁻¹) to the ferromagnetic coupling component that is initiated by the ion O4 located in the space of $J_1^{(2)}$ coupling, beside the O10 ion.

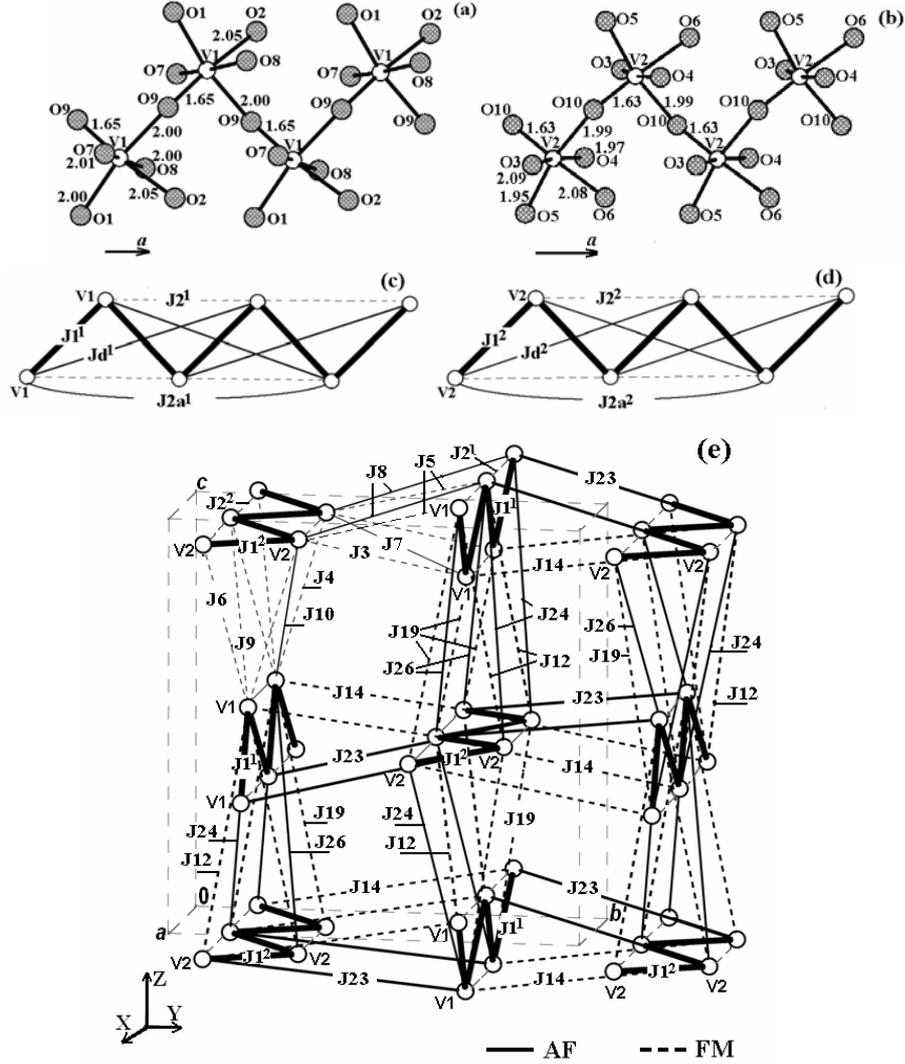

**Figure 2.** Two types [V(1)O₄O₂/₂]ₙ (a) and [V(2)O₄O₂/₂]ₙ (b) of zigzag chains of corner-sharing VO₆ octahedra running along the a-axis in ß-Sr(VOAsO₄)₂. The sublattice of V⁴⁺ and coupling $J_n$ in ß-Sr(VOAsO₄)₂: zigzag chains of V1 (c) and V2 (d) ions, (e) three-dimensional structure.

Next-to-nearest-neighbor couplings $J_2^{(1)}$ ($J_2^{s(1)}$ = 0.006 Å⁻¹ (FM), d(V1-V1) = 4.927 Å) and $J_2^{(2)}$ ($J_2^{s(2)}$ = 0.003 Å⁻¹ (FM), d(V1-V1) = 4.927 Å), unlike the nearest-neighbor couplings $J_1^{(1)}$ and $J_1^{(2)}$, are very weak FM ones. Therefore, there no competition between couplings $J_1^{(1)}$ and $J_2^{(1)}$ in the chain of V1 ions, as well as between couplings $J_1^{(2)}$ and $J_2^{(2)}$ in the chain of V2 ions.

Additional $J_{2a}^{(1)}$ and $J_{2a}^{(2)}$ couplings at long distances (d(V-V)=2$a$=9.854 Å) along the chain in parallel to the $a$ axis can not affect the state of zigzag chains, since they are very weak (figure 2(c), (d)). The coupling $J_{2a}^{(1)}$ ($J_{2a}^{s(1)}/J_1^{s(1)}$ = -0.013) in the chain of V1 ions is ferromagnetic, whereas the coupling $J_{2a}^{(2)}$ ($J_{2a}^{s(2)}/J_1^{s(2)}$ = -0.006) in the chain of V2 ions is, on the contrary, antiferromagnetic. However, both these couplings $J_{2a}^{(1)}$ and $J_{2a}^{(2)}$, as well as $J_2^{(1)}$ and $J_2^{(2)}$ couplings, are able to change their character even at insignificant changes of intermediate ions from weak FM to weak AF, including the cases of full absence of couplings. The instability of magnetic couplings in linear chains along the $a$ axis results from the fact that the sum of contributions into the FM component slightly exceeds the sum of these couplings contribution into the AF component. Beside the $J_1$ couplings, five times weaker additional AF- couplings $J_d^{(1)}$ ($J_d^{s(1)}/J_1^{s(1)}$ = 0.20, d(V1-V1) = 7.813 Å) and $J_d^{(2)}$ ($J_d^{s(2)}/J_1^{s(1)}$ = 0.21, d(V2-V2) = 7.802 Å) take place between these chains (figures 2(c), (d)).

Zigzag chains are coupled to each other relatively weakly (figure 2(e)). The strongest coupling in the V-V distance range up to 10 Å is the AF coupling $J_{23}$ ($J_{23}^s/J_1^{s(1)}$ = 0.21, d(V1-V2) = 8.344 Å) in the plane $ab$ between chains of ions V1 and chain of ions V2. Beside it, only two AF-couplings, $J_{30}$ (d(V1-V2) = 9.837 Å) and $J_{31}$ (d(V1-V2) = 9.888 Å), with the value $J_n^s/J_1^{s(1)} \geq 0.1$, take place in this plane. Antiferromagnetic couplings $J_{24}$ (d(V1-V2) = 8.418 Å) and $J_{26}$ (d(V1-V2) = 8.728 Å) between V1 and V2 chains in the plane $ac$ are approximately one and a half times weaker than $J_{23}$ ($J_n^s/J_1^{s(1)} \geq 0.13$-0.15), moreover, the $J_{26}$ coupling is unstable and can be reduced three-fold. The strongest ($|J_n^s/J_1^{s(1)}| = 0.15 - 0.18$) among the ferromagnetic interchain couplings are those $J_{12}$ (d(V1-V2) = 7.750 Å), $J_{14}$ (d(V1-V2) = 7.797 Å) and $J_{19}$ (d(V1-V2) = 7.918 Å), however, the latter two are unstable. The coupling $J_{14}$ can be reduced five-fold without changing the sign, while the coupling $J_{19}$ can transform into the AF state with $J_{19}^s/J_1^{s(1)} \sim 0.1$. The remaining interchain couplings of this compound are relatively weaker, including those at shorter distances V1-V2 in the range from 5.18 Å up to 6.45 Å, such as: AF $J_7$, $J_8$ и $J_{10}$, where the ratio $J_n^s/J_1^{s(1)}$ falls into the range from 0.03 up to 0.10, and FM $J_3 – J_6$ и $J_9$, where the ratio $|J_n^s/J_1^{s(1)}|$ falls into the range from 0.004 up to 0.02. One should mention that the strongest coupling among them $J_{10}$ can undergo the transition -0.015 Å$^{-1}$ (AF) → 0.010 Å$^{-1}$ (FM). The couplings $J_b^{(1)}$, $J_b^{(2)}$ (d(V-V) = 12.565 Å) and $J_c^{(1)}$, $J_c^{(2)}$ (d(V-V) = 12.739 Å) between the same vanadium ions located at the parameter distance of elementary cell along the axes $b$ and $c$ are weak ($J_n^s/J_1^{s(1)}$ = 0.08 – 0.14) antiferromagnetic ones.

Thus, the crystalline compound ß-Sr(VOAsO$_4$)$_2$ is S=1/2 one-dimensional antiferromagnet with alternating along the axes $b$ and $c$ zigzag spin chains of ions V1 and V2 running along the $a$ axis. Zigzag spin chains in this compound shall be considered as single chains, according to [13-16], since the values of ratios of AF nearest- and next-to-nearest-neighbor couplings $J_2/J_1$ are very small ($|J_2^{s(1)}/J_1^{s(1)}|$ = 0.04 and $|J_2^{s(2)}/J_1^{s(2)}|$ = 0.02). The interchain couplings $J_3$ - $J_6$ at short distances (d(V1-V2) = 5.185-5.495 Å) are weak FM couplings ($|J_n/J_1|$ = 0.004-0.020). Stronger interchain couplings $J_{23}$ ($J_{23}^s/J_1^{s(1)}$ = 0.21) emerge at long distances (d(V1-V2) = 8.344 Å) in the plane $ab$, these couplings are too five times weaker than intrachain nearest-neighbor ones.

### 3.3. (NH$_4$)$_2$VOF$_4$ и K$_2$VOF$_4$

The isostructural compounds (NH$_4$)$_2$VOF$_4$ [7] and K$_2$VOF$_4$ [8] crystallize in the Pna2$_1$ system with a= 7.596(7.403) Å, b= 12.043(11.443) Å, and c= 5.770(5.650) Å. One should mention that in [8] the space group for the K-compound is given in the alternative setting Pn2₁a, but we will consider the unit cell $K_2VOF_4$ in the standard setting. The positions of ions F1, F2, F3, O and F4 in the $NH_4$-compound correspond to the positions F3, F2, F1, 1/2O1-1/2F4 and 1/2O2-1/2F5 in the K-compound. All the data for $K_2VOF_4$ corresponding to the data for $(NH_4)_2VOF_4$, will be further given in round brackets.

The $V^{4+}$ ion is surrounded by five fluorine atoms (d(V-F)=1.910-2.224(1.868-2.163) Å) and one oxygen atom (d(V-O)= 1.612 Å; in $K_2VOF$ d(V-O/F)= 1.703 and 1.783 Å), forming the $VF_5O$ octahedron. The bond valence sums of V (3.94(3.87)) are in a good agreement with the expected values. These octahedra share corners F3(F1) (d(V-F3)=1.982(2.043) Å and 2.224(2.163) Å) forming zigzag chains parallel to the $c$ axis (figure 3(a)). In the zigzag chain the V-F3-V bond angles are equal to 172(171)°.

Magnetic ions $V^{4+}$ form zigzag (angles V-V-V are equal to 86.9(84.7)°) chains along the $c$ axis. The magnetic and crystalline structures of the magnetic ions sublattice coincide, as in the compound ß-Sr(VOAsO₄)₂ (figure 3).

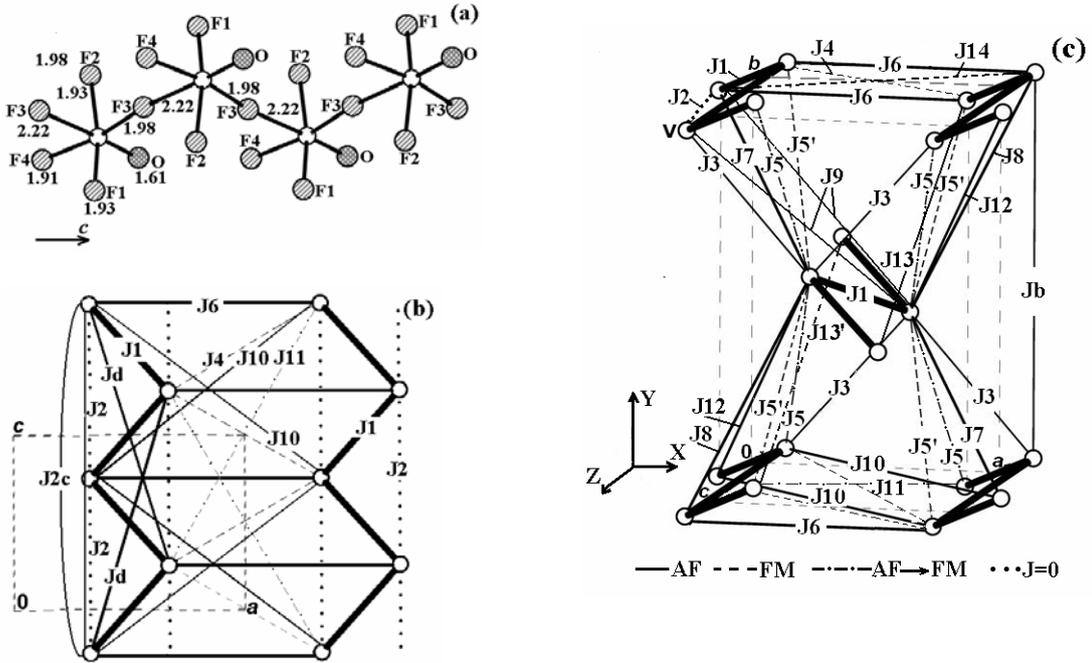

**Figure 3.** Zigzag chains of corner-sharing $VO_6$ octahedra in $(NH_4)_2VOF_4$ (a) and sublattice of $V^{4+}$ and coupling $J_n$: (b) $ac$ plane, (c) three-dimensional structure.

The intrachain nearest-neighbor couplings $J_1$ ($J_1^s$ = -0.135(-0.129) Å⁻¹, d(V-V)=4.196(4.193) Å) are dominating antiferromagnetic ones (figure 3(b), (c)). The emerging of this strong AF- couplings is initiated by the F3 ions, since they are located near the centre of the straight line V-V, coupling the vanadium ions ($l'/l$ = 1.11(1.06) and $h$(F3) = 0.147(0.168) Å). The next-to-nearest-neighbors $J_2$ couplings ($J_2^s$ = 0(0), d(V-V) = 5.770(5.650) Å = $c$) are absent, since small contributions into AF (from F4 and O1) and FM (from two F3 ions) components of coupling are approximately equal and suppress each other. However, insignificant displacements of these intermediate ions located in critical positions (critical point "b"; see section 3 in [1]) can result in emerging of small couplings of both AF and FM type. It is interesting that weak antiferromagnetic couplings $J_{2c}$ ($J_{2c}^s / J_1^s$ = 0.07(0.01)) at long distances (d(V-V) = 11.540(11.300) Å = $2c$) exist along the zigzag chain parallel to the $c$ axis (figure 3(b)). In determining the character of these couplings the primary importance belongs to the AF-contribution ($j_V$ = -0.009(-0.009) Å$^{-1}$) initiated by the V$^{4+}$ ion. In addition to the $J_1$ couplings between linear chains, there exist AF $J_d$ ($J_d^s/J_1^s$ = 0.24(0.25), d(V-V) = 9.176(9.023) Å) couplings that form AF-triangles with $J_1$ и $J_{2c}$, in which competition is highly unlikely, since one of the interactions ($J_{2c}$) is too weak.

The strongest among the interchain couplings in the NH$_4$-compound are those AF $J_6$ ($J_6^s/J_1^s$ = 0.25, d(V-V) = 7.596 Å = $a$) along the $a$ axis in planes parallel to the $ac$ plane and $J_7$ ($J_7^s/J_1^s$ = 0.23, d(V-V) = 8.193 Å) and $J_8$ ($J_8^s/J_1^s$ = 0.24, d(V-V) = 8.530 Å) between these planes (figure 3(b), (c)). Main contribution to the AF components of the $J_6$ and $J_7$, $J_8$ couplings emerges under effect of intermediate N2 and N1 ions, respectively. In the K-compound these interchain couplings appeared to be weak ($J_6^s/J_1^s$ = -0.03, d(V-V) = 7.403 Å, $J_7^s/J_1^s$ = 0.02, d(V-V) = 7.958 Å, $J_8^s/J_1^s$ = 0.02, d(V-V) = 8.126 Å), whereas $J_6$ were ferromagnetic while $J_7$ and $J_8$ - antiferromagnetic, since during calculations of the sign and strength of these couplings K$^{1+}$ were not taken into account. (The studies conducted in [1] have brought us to the conclusion that metal cations without unpaired electrons do not participate in the magnetic couplings formation.) The strongest interchain couplings in the K-compound are those AF $J_{10}$ and $J_{10'}$ ($J_{10}^s/J_1^s$ = 0.21, $J_{10'}^s/J_1^s$ = 0.18, d(V-V) = 9.313 Å) in the $ac$ plane (figure 3(b), (c)). Maximum couplings $J_{12}$ and $J_{12'}$ ($J_{12}^s/J_1^s$ = 0.16, $J_{12'}^s/J_1^s$ = 0.15, d(V-V) = 9.897 Å) between planes are markedly weaker than $J_{10}$ and $J_{10'}$ couplings in the $ac$ plane. Parameters of these couplings in the NH$_4$-compound are about the same as in the K-compound ($J_{10}^s/J_1^s$ = 0.21, $J_{10'}^s/J_1^s$ = 0.17, d(V-V) = 9.539 Å; $J_{12}^s/J_1^s$ = 0.18, $J_{12'}^s/J_1^s$ = 0.17, d(V-V) = 10.298 Å).

Changing of the type of ordering at substitution of NH$_4$ by K occurs, aside from $J_6$, in three other weak couplings $J_4$, $J_{13}$ and $J_b$. The interchain $J_4$ coupling in the $ac$ plane in the NH$_4$-compound are ferromagnetic ($J_4^s/J_1^s$ = -0.004, d(V-V) = 6.006 Å), while in the K-compound, on the contrary, antiferromagnetic ($J_4^s/J_1^s$ = 0.005, d(V-V) = 5.688 Å). However, insignificant displacements of intermediate ions of oxygen fluorine can result as in its complete disappearance as in the transition AF-FM, since the sum of contributions into AF and FM components of the $J_4$ coupling are virtually equal. Antiferromagnetic character of ordering of the $J_{13}$ coupling ($J_{13}^s/J_1^s$ = 0.09, d(V-V) = 10.613 Å) in the NH$_4$ - compound is initiated by the intermediate ion N2, which contribution into the AF component of coupling significantly exceeds the sum of contributions from intermediate fluorine ions into the FM-component of this coupling. In the K-compound, in which there is no such a contribution, the $J_{13}$ coupling is very weak and ferromagnetic ($J_{13}^s/J_1^s$ = -0.04, d(V-V) = 10.274 Å). It occurs similarly for the $J_b$ coupling (figure 3(c)), which is antiferromagnetic and stronger in the NH$_4$ -compound ($J_b^s/J_1^s$ = 0.12, d(V-V) = 12.043 Å) and ferromagnetic and very weak in the K-compound ($J_b^s/J_1^s$ = -0.02, d(V-V) = 11.443 Å).

The remaining interchain couplings in the NH$_4$-compound are comparable with respective couplings in the K-compound [$J_3^s/J_1^s$ = 0.12(0.15), d(V-V) = 5.817(5.605) Å; $J_5^s/J_1^s$ = 0.03(0.02), $J_{5'}^s/J_1^s$ = -0.11(-0.07), d(V-V) = 6.787(6.458) Å; $J_9^s/J_1^s$ = 0.09(0.09), $J_{9'}^s/J_1^s$ = 0.13(0.10), d(V-V) = 9.235(9.041) Å; $J_{11}^s/J_1^s$ = -0.16(-0.11), d(V-V) = 10.132(9.808) Å; $J_{13'}^s/J_1^s$ = -0.04(-0.07), d(V-V) = 10.613(10.274) Å; $J_{14}^s/J_1^s$ = -0.15(-0.013), d(V-V) = 10.702(10.602) Å].

Thus, we have shown that isostructural compounds (NH$_4$)$_2$VOF$_4$ and K$_2$VOF$_4$ are 1-D antiferromagnets with zigzag spin-1/2 chains along $c$ axis. According to [13-16], the zigzag chains in (NH$_4$)$_2$VOF$_4$ и K$_2$(NH$_4$)$_2$VOF$_4$ are single Heisenberg chains, since $J_2$ = 0. The maximum strength of interchain couplings is weaker than that of intrachain ones in four times in (NH$_4$)$_2$VOF$_4$ ($J_6^s/J_1^s = 0.25$) and in five times in $K_2VOF_4$ ($J_{10}^s/J_1^s = 0.21$). Here the maximum strength of interchain couplings in the compound $(NH_4)_2VOF_4$ is the same as in planes parallel to the *ac* plane as between them, whereas in the compound $K_2VOF_4$ the coupling within the planes is 1.4 times stronger ($J_{10}^s/J_{12}^s = 1.36$) than between planes.

### 3.4. α-ZnV₃O₈

α-ZnV₃O₈ [9] crystallized in the orthorhombic system with space group Iba2 and cell parameters $a$ = 14.298(5) Å, $b$ = 9.908(3) Å, $c$ = 8.430(3) Å. In the crystal structure of α-ZnV₃O₈ one can mark out four types of metal sites differing by the coordination surrounding, namely, trigonal bipyramidal V1 (V1-O = 1.59 – 2.07 Å), octahedral V2 (V2-O = 1.68-2.018), tetrahedral V3 (V3-O = 1.69-1.80), and octahedral Zn (Zn-O = 1.91-2.018). The valent states of metallic ions were estimated as V1(V), V2(IV), V3(V) and Zn(II), whereas the bond-valence sums [12] of V and Zn ions have the following values: 5.15 for V1, 3.62 for V2, 5.0 for V3 and 2.39 for Zn. However, in spite of the characteristic for the V⁴⁺ ion distortion of the octahedron V2 (the vanadyl bond V2-O6 = 1.681 Å), one may suggest that there exists a partial disordering of the V⁴⁺ and Zn²⁺ ions in the positions V2 and Zn, since the BVS values of V2 and Zn deviate substantially from ideal values. According to the BVS values, the positions V2 must be occupied by 81% V⁴⁺ and 19% Zn, whereas the sites Zn, on the contrary, by 81% Zn and 19% V⁴⁺.

One should mention that the ICSD database cites the oxidation number of the vanadium ion in the position V1 mistakenly as 4 and in the position V2 as 5. Our preliminary calculations, based on the assumption that the magnetic ion is in the position V1 (not V2) showed that this compound is S = 1/2 spin-dimer antiferromagnet (intradimer coupling: $J_1^s = -0.088$ Å¹, d(V1-V1) = 3.347 Å; maximum interdimer coupling: $J_{12}^s = -0.036$ Å⁻¹, d(V1-V1) = 9.229 Å) that is similar to a distorted SrCu₂(BO₃)₂ [21].

Let us assume that there exists an ordering between V2 and Zn ions and consider the magnetic structure of the compound α-ZnV₃O₈ formed by magnetic couplings between V⁴⁺ ions located in the position V2. In the crystal structure of α-ZnV₃O₈ octahedra of magnetic ions V2O₆ do not share oxygen atoms. The V2 ions located at shortest distances 5.312 Å ($J_1$-coupling) and 5.347 Å ($J_2$-coupling) form corrugated planes perpendicular to the *a* axis from zigzag chains running along the *c* axis. The shortest distance between these planes is 5.476 Å ($J_3$-coupling) (figure 4 (a)). However, estimation of the spin-spin interactions shows that $J_1$, $J_2$ and $J_3$ couplings between three closest neighbouring V2 ions are very weak.

The fourth-nearest-neighbor AF $J_4$ coupling ($J_4^s = -0.070$ Å⁻¹, d(V2-V2) = 5.676 Å) between V2 ions (∠V2V2V2 = 121.56°) zigzag-like chains arranged along the *b* axis appeared to be the dominating coupling (figure 4). The main contribution (0.079 Å⁻¹) into the AF-component of the $J_4$ coupling emerges under effect of the O2 ion located virtually on the straight line V2-V2 (distance $h$(O2) from the centre of the O2 ion to the straight line is equal to 0.268 Å) in the central one-third part of space between magnetic ions ($l'/l$ = 1.62) (figures 4(b), 5(a)). Contributions from ions O4 ($j_{O4}$ = 0.003 and 0.002 Å⁻¹), O7 ($j_{O4}$ = 0.002 Å⁻¹), O1($j_{O1}$ = 0.002 Å⁻¹) and O6 ($j_{O6}$ = -0.0003 Å⁻¹) into FM and AF components of this coupling are insignificant. The next-to-nearest-neighbor couplings $J_{12}$ ($J_{12}^s = 0.001$ Å⁻¹, d(V2-V2) = 9.908 Å = $b$) in these zigzag chains (figure 4(c)) are very weak and ferromagnetic. Contributions initiated by every of nine intermediate oxygen atoms into the AF and FM components of the $J_{12}$ coupling are small while their sums are virtually equal. Besides, insignificant displacements of these ions can result as in elimination of this coupling as in its transition into weak AF state.

The zigzag chains are bound by strong AF $J_6$ couplings ($J_6^s/J_4^s = 0.46$, d(V2-V2) = 7.340 Å) into planes parallel to the *ab* plane (figure 4(c), (d)). The coupling $J_6$ under effect of the intermediate ion O5 from the trigonal bipyramid V1 (figure 5(b)). Besides, relatively strong FM $J_8$ couplings ($J_8^s/J_4^s$ = -0.33, d(V2-V2) = 7.936 Å) and weaker AF $J_3$ ($J_3^s/J_4^s$ = 0.15, d(V2-V2) = 5.476 Å) and FM $J_{15}$ ($J_{15}^s/J_4^s$ = -0.16, d(V2-V2) = 10.451 Å) couplings take place between chains in these planes. However, the strength of the FM $J_{15}$ coupling can increase more than twice (up to $J_{15}^s/J_4^s$ = -0.40) in the case of insignificant displacement (up to $l_n/l_n < 2.0$) of intermediate ions O₅ localized in the critical point "c" (see section 3 in [1]). One should emphasize that intrachain $J_4$ and interchain $J_3$ and $J_6$ couplings that form AF triangles in the plane *ab* (figure 4(c)) compete with each other. However, the degree of frustration of these interactions is hard to estimate, since one of three couplings ($J_3$) in the triangle is much weaker than other two ($J_3^s/J_4^s$ = 0.15, $J_6^s/J_4^s$ = 0.46).

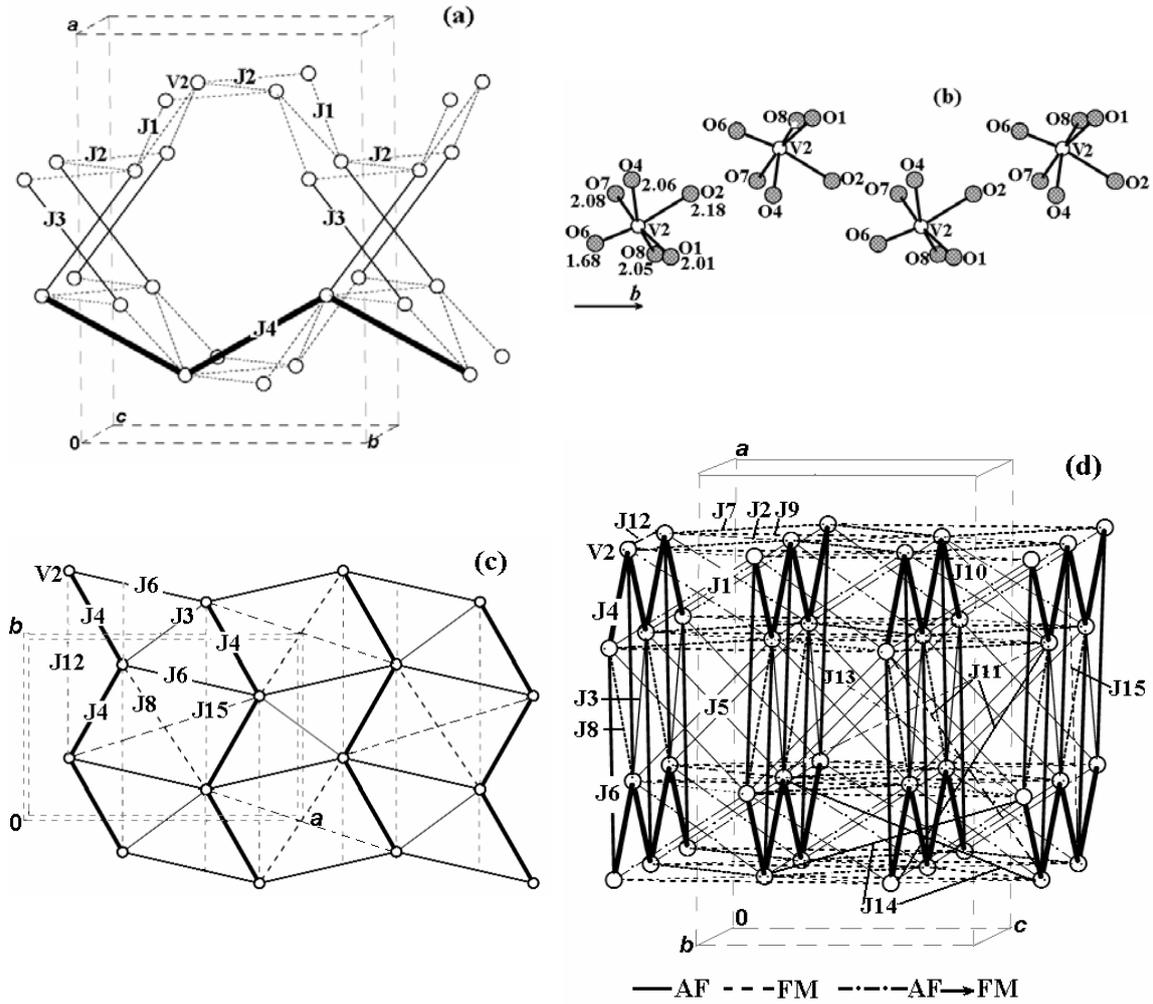

**Figure 4.** α-ZnV₃O₈: (a) Zigzag chains of V2 ions along the *c* axis ($J_1$- and $J_2$-couplings) and *b* axis ($J_4$-couplings); (b) the zigzag arrangement of the octahedra V2 along the *b*-axis; the sublattice of V2 and coupling $J_n$: (c) *ab* plane and (d) three-dimensional structure.

The planes with strong interchain couplings draw up at short distances from each other perpendicular to the *c* axis (figure 4(d)). Weak FM couplings $J_1$ ($J_1^s/J_4^s$ = -0.05) и $J_2$ ($J_2^s/J_4^s$ = -0.01) mentioned earlier appeared to be interplane couplings. One should also mention that the $J_1$-coupling is unstable, since there is a possibility of transition from FM ($J_1^s = 0.004$ Å⁻¹) into AF ($J_1^s = -0.019$ Å⁻¹) state due to five-fold increase of the contribution initiated by the ion O(4) into the AF coupling component in case of its insignificant (up to $l_n'/l_n < 2.0$) displacement from the critical point "c" (figure 5(c)). The AF $J_5$ ($J_5^s/J_4^s = 0.07$, d(V2-V2) = 6.077 Å), $J_{10}$ ($J_{10}^s/J_4^s = 0.04$, d(V2-V2) = 9.664 Å) and FM $J_{13}$ ($J_{13}^s/J_4^s = -0.03$, d(V2-V2) = 10.052 Å) couplings and unequal $J_{11}$ (d(V2-V2) = 9.665 Å) couplings (among the latter, two couplings are AF ones ($J_{11}^s/J_4^s = 0.13$, $J_{11'}^s/J_4^s = 0.04$) while two other - FM ones ($J_{11''}^s/J_4^s = -0.09$, $J_{11'''}^s/J_4^s = -0.04$)) are the same weak couplings between chains from nearest planes.

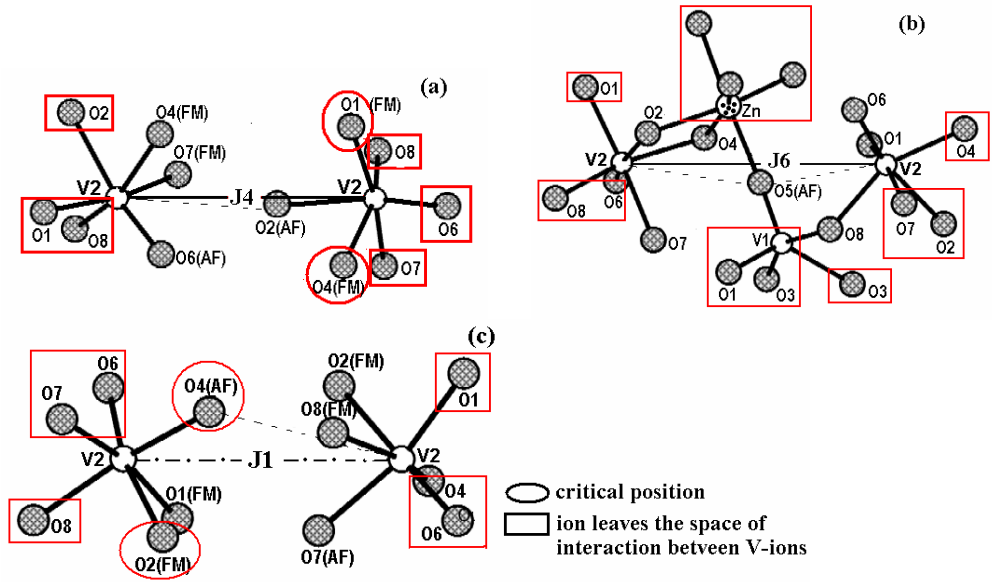

**Fig. 5.** The arrangement of the intermediate ion $O_n$ in space between V2 ion: (a) $J_4$, (b) $J_6$ and $J_1$ couplings.

The FM $J_7$ coupling ($J_7^s/J_4^s = -0.34$, d(V2-V2) = 7.847 Å) between neighbouring planes, the FM $J_9$ ($J_9^s/J_4^s = -0.33$, d(V2-V2) = 8.430 Å = $c$) and AF $J_{14}$ (d(V2-V2) = 10.163 Å) couplings between next-to-nearest neighboring planes are relatively strong interplane ones, but weaker than the $J_6$-coupling inside the plane. However, the $J_{14}$ couplings are unequal, since one of them ($J_{14}$) is strong ($J_{14}^s/J_4^s = 0.40$), whereas another one ($J_{14'}$) is weak ($J_{14'}^s/J_4^s = 0.08$). Besides, the $J_{14}$ coupling can strengthen in 1.4 times at insignificant displacement (by 0.02 Å) of the intermediate ion O2 from the line V-V and its exit beyond the coupling space.

Let us discuss a hypothetical variant when in the structure $\alpha$-ZnV$_3$O$_8$ magnetic ions with radius equal to the radius V⁴⁺ (0.58 Å) occupy the positions of Zn and V2 ions. Let us mark the magnetic ions localized in the Zn position as Z and those in the V2 position as V2, (as before). In this case the crystalline structure of the magnetic ions sublattice would comprise zigzag chains along the $b$ axis made of shared triangle edges (figure 6). According to our calculations, $J_{4(VV)}$ coupling (corresponds to $J_4$ calculated above) between V2 ions in zigzag-like chains along the $b$ axis would be frustrated. Frustration would originate from the competition between strong AF couplings in triangles of two types.

One triangle is formed by the AF $J_{3(ZV)}$ ($J_{3(ZV)}^s = -0.132$ Å⁻¹, d(Zn-V2) = 3.462 Å), $J_{4(VV)}$ ($J_{4(VV)}^s/J_{3(ZV)}^s = 0.53$, d(V2-V2) = 5.676 Å) and $J_{2(ZV)}$ ($J_{2(ZV)}^s/J_{3(ZV)}^s = 0.32$, d(Zn-V2) = 3.370

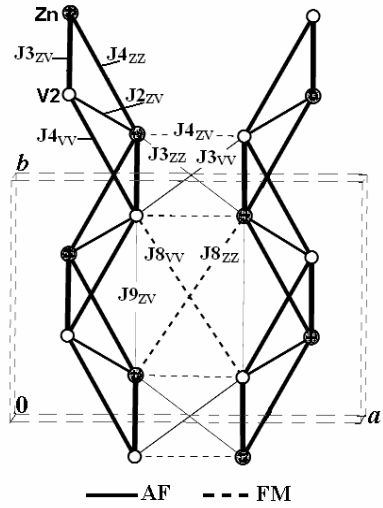

**Fig. 6.** The $J_n$ coupling in hypothetical chains formed by magnetic ions in positions Zn and V2.

Å) couplings, whereas another one - by the $J_{3(ZV)}$, $J_{4(ZZ)}$ ($J_{4(ZZ)}^s / J_{3(ZV)}^s = 0.58$, d(Zn-Zn) = 5.671 Å) and $J_{2(ZV)}$ couplings. Between these hypothetical magnetic chains, the following couplings take place: strong nearest-neighbor AF $J_1$ couplings between ions in positions V2 and Zn ($J_{1(ZV)}^s / J_{3(ZV)}^s = 0.35$, d(Zn-V2) = 3.207 Å); very weak AF couplings $J_{3(VV)}$ ($J_{3(VV)}^s / J_{3(ZV)}^s = 0.08$, d(V2-V2) = 5.476 Å) and $J_{9(ZV)}$ ($J_{9(ZV)}^s / J_{3(ZV)}^s = 0.0003$, d(Zn-V2) = 6.672 Å); FM couplings $J_{4(ZV)}$ ($J_{4(ZV)}^s / J_{3(ZV)}^s = -0.13$, d(Zn-V2) = 4.498 Å), $J_{8(ZZ)}$ ($J_{8(ZZ)}^s / J_{3(ZV)}^s = -0.30$, d(Zn-V2) = 7.904 Å) and $J_{8(VV)}$ ($J_{8(VV)}^s / J_{3(ZV)}^s = -0.17$, d(V2-V2) = 7.936 Å).

Thus, our calculations show that in case of ordering $V^{4+}$ and $Zn^{2+}$ ions distribution on crystallographic sites the magnetic structure of α-ZnV$_3$O$_8$ would consist of antiferromagnetic S=1/2 zigzag chains. These chains form due to the $J_4$ ($J_4^s$ = -0.070 Å⁻¹, d(V2-V2) = 5.676 Å) couplings between zigzag-like arranged along the *b* axis $V^{4+}$ ions in the position V2 and comprise, according to [13-16], single chains, since the value of the ratio of the nearest- ($J_4$) and next-to-nearest-neighbor ($J_{12}$) couplings is very small ($|J_{12}^s / J_4^s| = 0.01$). The chains are bound to each other into planes parallel to the *ab* plane by strong AF couplings ($J_6^s / J_4^s = 0.46$). Between these planes, interchain AF ($J_{14}^s / J_4^s = 0.40$) and FM ($J_7^s / J_4^s = -0.34$, $J_8^s / J_4^s = -0.33$, $J_9^s / J_4^s = -0.33$) couplings are little bit weaker. Competition takes place between unequal AF intrachain $J_4$ and interchain $J_3$ and $J_6$ couplings ($J_6^s / J_4^s = 0.46$, $J_3^s / J_4^s = 0.15$), since they form a triangle. At partial disordering of the ions $V^{4+}$ and $Zn^{2+}$ in the positions V2 and Zn, confirmed by calculations of BVS of V2 and Zn, frustration of all strong AF couplings in the compound α-ZnV$_3$O$_8$ could occur.

## 4. Conclusions

Crystalline structure is very important in formation of the insulators magnetic structure. The existence of a specific type of magnetic structure and the presence of competition between couplings is determined by spatial location of magnetic ions in a crystal in combination with the magnetic couplings parameters. The magnetic couplings parameters (strength of magnetic couplings and type of spin ordering) are determined by the size and geometrical location of intermediate ions in local space between magnetic ions and also by distances between magnetic ions. In [1] we developed the method for calculation of magnetic couplings parameters on the basis of structural data, whereas in this work we demonstrated its application in search of spin-1/2 zigzag chain antiferromagnets.

Five AF zigzag spin chain compounds - $In_2VO_5$, ß-$Sr(VOAsO_4)_2$, $(NH_4)_2VOF_4$, $K_2VOF_4$ and $\alpha$-$ZnV_3O_8$ – were found and studied. The magnetic structures of all the compounds, except $\alpha$-$ZnV_3O_8$, correspond to the crystal structure of the $V^{4+}$ magnetic ions sublattice. In the crystal $\alpha$-$ZnV_3O_8$ one can select several chains (figure 4(a)) from zigzag-like arranged magnetic ions V2, however, only one of them (that with the longest distances V2-V2) would have strong AF couplings due to respective location of intermediate ions.

The calculations have shown that three compounds - ß-$Sr(VOAsO_4)_2$, $(NH_4)_2VOF_4$ and $K_2VOF_4$ – are magnetically ordered S=1/2 one-dimensional antiferromagnets with zigzag spin chains comprising single chains. The strongest interchain couplings in these couplings are 4-5 weaker than intrachain nearest-neighbor couplings. The magnetic structures of two other compounds are much more complicated and are of interest for studies of magnetic frustrations. Antiferromagnetic zigzag spin chains in the compound $\alpha$-$ZnV_3O_8$ are also single chains. However, relatively strong interchain couplings take place between these chains. The nearest-neighbor intrachain coupling competes with two unequal on strength interchain couplings forming together a triangle. Besides, in case of partial disordering of $V^{4+}$ and $Zn^{2+}$ ions in the positions V2 and Zn, confirmed by calculations of BVS of V2 and Zn, zigzag chains transform into frustrated antiferromagnetic triangles located along the chain. The compound $In_2VO_5$ zigzag spin chains are frustrated, since the ratio of competing AF nearest- ($J_1$) and AF next-to-nearest-neighbor ($J_2$) couplings is equal to $J_2/J_1 = 1.68$ that exceeds the Majumdar-Ghosh point by 1/2. Between zigzag chains in the *ab* plane, relatively strong AF couplings take place - they are two times stronger than maximum couplings between chains from neighbouring planes.

## Acknowledgment

This work is supported by grant 06-I-P8-009 of the Far Eastern Branch of the Russian Academy of Sciences.